# An Adaptive Load Balancing to Provide Quality of Service

[1] Zahra Vali, [2] Massoud Reza Hashemi, [3] Neda Moghim
[*1,] Isfahan University of Technology, Isfahan, Iran
[2,] Isfahan University of Technology, Isfahan, Iran
[3,] University of Isfahan, Isfahan, Iran
Email: [1]z.vali@ec.iut.ac.ir, [2]hashemim@cc.iut.ac.ir, [3]n.moghim@eng.ui.ac.ir

*Abstract.* **In order to utilize network resources and provide a more reliable delivery of information, multipath routing algorithms are used with a higher priority than single path routing. But providing the quality of service (QoS) requirements while load balancing is done among multiple paths is still a challenge. In this paper a new load balancing algorithm is proposed by using an explicit endpoint admission control (EEAC) to achieve a dynamic load balancing algorithm, with the ability to guarantee the end-to-end QoS for a variety of service classes. The simulation results for nondeterministic network conditions show that the proposed algorithm increases the utilization of network resources and also decreases the end-to-end delay.**

**Keywords**: multipath routing, load balancing, quality of service, admission control

* Corresponding Author:
Zahra Vali,
Faculty of Electrical and Computer Engineering,
Isfahan University of Technology Isfahan, Iran,
Email: z.vali@ec.iut.ac.ir    Tel:+989132677304

## 1. Introduction

In recent years, Internet has been changed as a way to provide different services with different quality of service (QoS) guarantees in response to variable users' applications. Traffic engineering (TE) has been created to improve network performance by providing the QoS requirements of end users and decreasing network congestion. Among different TE approaches, routing algorithms play a key role in providing a desirable performance in the network.

In order to utilize network resources and improve the reliability of delivering information, multipath routing algorithms are used with higher priority than single path routing. With the Cooperation of load balancing algorithms, multipath routing can lead to higher bandwidth efficiency and also lower congestion in the network. According to optimized routing, finding multiple efficient paths and balancing the network load on them while QoS requirements are also provided, still need more research works [1].

According to the classification of TE approaches in [2], load balancing algorithms belong to state-dependent TE approaches that are concerned with relatively fast network state changes while time-dependent algorithms are used for utilization of network resources in response to long timescale traffic changes. Some requirements of load balancing mechanisms are explained in [3]. Load balancing mechanisms can be divided into two groups according to the dependency to the network states: dependent and independent load balancing algorithms. Some independent load balancing methods have been stated in [4, 5] as below:
-Per packet traffic splitting in a round robin order
-Splitting destination prefix into available next hops

-Traffic splitting according to a hash function applied to the header fields of the data packets

In the independent load balancing algorithms, traffic splitting is done with no consideration of network congestion states and the possibility of congestion cannot be omitted. State dependent algorithms perform traffic splitting regarding the network states, so they are more effective solutions to reduce the possibility of network congestion.

To perform dynamic load balancing, first the appropriate multiple paths should be found with the use of an appropriate multipath routing algorithm [6]. The next step is how to collect and advertise the information of network congestion. One way is to send a stream of probing packets from the source to the destination. Then, an algorithm is required to calculate the proper traffic rate for each of the available path based on the collected information of the received probing packets. This algorithm is generally based on optimization or heuristic functions.

Despite the accuracy of the optimization-based functions, imposing a high complexity to the network elements turns them to inapplicable solutions for large networks. Slow convergence is another problem associated with the optimization-based functions. In contrast, heuristic functions which are using the information of probing packet trains and simple equations, are more applicable, more scalable and less expensive to be implemented in large networks.

In [7, 8, 9], state-dependent algorithms based on the heuristic functions which use the probing packets to measure the average one-way delay are introduced. These algorithms cannot lead to a proper dynamic load balancing because the result of the measurement is just related to an instant sampling of highly dynamic network congestion state.

The proposed algorithm is a state-dependent algorithm that does not suffer from the above problem. In this algorithm, probing packets are sent periodically in the network and information related to network congestion in the probing packets is updated in each probing phase.

The organization of this paper is as follows. The proposed algorithm is introduced in section 2. Simulation scenarios and network topology is explained in section 3 and section 4 contains the conclusion.

## 2. Proposed algorithm

The proposed algorithm is based on an explicit endpoint admission control algorithm to provide necessary information for balancing load among multiple paths [10]. Furthermore, to guarantee the end-to-end QoS requirements for all service classes in the proposed algorithm, explicit endpoint admission control algorithm helps differentiated service (DiffServ) to accept traffic demands and simultaneously a load balancing algorithm divides the accepted traffic among multiple paths for higher network utilization. The importance of the combination of endpoint admission control algorithm and the proposed load balancing technique is that, the end point admission control provides the required information of network states dynamically for load balancing algorithm [11]. On the other hand, explicit endpoint admission control offers a higher variety of service classes compared to DiffServ architecture [12] by using service vectors [11].

The explicit endpoint admission control algorithm consists of probing and data transfer phases. In the probing phase, probing packets are sent through the network to collect the required information for a new user request. Probing phase is repeated every T seconds to adapt the network dynamism with new users' requests.

Multiple path algorithm (MPA) is used to find multiple paths in the probing phase [13]. As the probing packet passes the routers along the paths, each router checks the availability of the requested service class. If the requested service class is not available, lower service classes will be checked and finally the available service class in the router will be identified and its availability will be written as a binary code in a two bits field in the probing packet [10]. Three classes of service are considered: EF, AF and BE which are equal to 00, 01 and 10 binary code respectively [14].

After analyzing the information of the probing packets at the destination, two, or generally more, paths are selected among all probed paths. According to the previous studies, routing on two paths is significantly different from single path routing but routing on more than two paths is not very different from routing on two paths [8]. Therefore two paths are selected at the destination to simplify the algorithm.

## 2.1. Evaluation of service classes

When a probing packet passes through the path, the availability of service classes at each router should be determined. In order to avoid oscillation which can be caused by bursty traffics, exponentially moving average (EMA) equations are used to average the incoming data rate and buffer length in each router for all classes [14]. All calculations of EMA equation are done in the routers and there is no need to be kept in the probing packets.

$$\bar{r}_{sj}(t) = \left(1 - e^{\frac{-\tau_{sj}(t)}{k}}\right) \times \frac{l_{sj}(t)}{\tau_{sj}(t)} + e^{\frac{-\tau_{sj}(t)}{k}} \times \bar{r}_{sj,old}(t) \quad (1)$$

$sj$: service class

$\bar{r}_{sj}(t)$ : The average arrival data rate

$l_{sj}(t)$ : The length of the received packet at time t

$\tau_{sj}(t)$ : The period between the current time t and the reception time of the previous packet

$\bar{r}_{sj,old}(t)$ : The last updated average data rate

k : a constant value (0.01)

$$\bar{L}_{sj}(t) = \left(1 - e^{\frac{-\tau_{sj}(t)}{k}}\right) \times L_{sj}(t) + e^{\frac{-\tau_{sj}(t)}{k}} \times \bar{L}_{sj,old}(t) \quad (2)$$

$\bar{L}_{sj}(t)$ : The average of buffer length

$\bar{L}_{sj,old}(t)$ : The last updated average buffer length

$L_{sj}(t)$ : Buffer length at time t

Since the measurement of $\bar{r}_{sj}(t)$ and $\bar{L}_{sj}(t)$ are done in each packet arrival with the cooperation of routers along the path, the obtained performance of service classes are the reflection of real network state over the time. In the proposed algorithm, $\bar{r}_{sj}(t)$ and $\bar{L}_{sj}(t)$ are three dimensional arrays that are updated according to the arrival of three service classes.

EMA equations consist of two parts: one part is related to current observation and the other part is related to past observations. If current observations are much bigger than previous ones, the equations cannot filter the transient effect of bursty traffics. On the other hand, if previous observations are much bigger than current ones, EMA equations converge slowly to the real value of traffic. k is a constant that controls the behavior of the bursty traffics. The calculation of the acceptable bounds of k in a similar algorithm is introduced in [10].

It is assumed that buffer lengths of different service classes are not the same in each router but buffer lengths of similar classes are the same in all routers.

Weighted fair queuing (WFQ) is used as the packet scheduling algorithm while other algorithms are also acceptable. WFQ can guarantee delay limitation and fairly assign resources among different service classes. The incoming traffic rate is compared to the maximum allowable bandwidth allocated to the requested service class in WFQ. If the arrival rate is lower than the allocated rate, then the traffic will be accepted otherwise the lower service classes are considered.

### 2.2. Decision function and load balancing parameter

End-to-end delay or packet loss probability alone cannot provide a good estimation of real network congestion state for traffic splitting [9]. Buffer length is used in this paper to estimate the congestion state of the paths instead of both end-to-end delay and packet loss probability. Remaining buffer space for each class of service is calculated as follows:

$$B_{sj} = L_{sj} - \bar{L}_{sj}(t)$$

$L_{sj}$ : The maximum buffer length of service class $s_j$

As the probing packet goes through the path, the remaining buffer spaces of all classes in each router are compared to the values of previous router and the bottleneck of the network is detected and inserted in the probing packets. After obtaining the information of each probing packet at the destination, remaining buffer space of the three service classes of the most crowded router in all path are added together.

$$Sum(i) = B_{EF}(i) + B_{AF}(i) + B_{BE}(i)$$

$Sum(i)$: Summation of the remaining buffer space of three service classes in the most crowded router of path i

$Sum(i)$ is chosen as a criterion to select the most proper paths; two paths with the most amount of $Sum(i)$ are selected. Also it is used as a limiting factor for sending traffic through the network and the incoming traffic will be divided on the two previously selected paths according to the rates calculated in equation (5) and a KP producing uniform random numbers as a traffic splitter.

$$Rate\ Path\ (i) = \frac{Sum(i)}{\sum_i Sum\ (i)}$$

### 3. Simulation results

The proposed algorithm is implemented in a network topology shown in Fig. 1 that is composed of one pair of source-destination, seven routers and 4.5 Mbps connecting link. OPNET modeler is used for the evaluation of the proposed algorithm [15]. Equation (1) and (2) are updated each time a probing or data packet is received. These two equations are written in all routers with state variables arrays indicating $\bar{r}_{sj}(t)$ and $\bar{L}_{sj}(t)$ for three service classes. After each update, the current values of $\bar{r}_{sj}(t)$ and $\bar{L}_{sj}(t)$, are stored as old variables for the next round of calculations. Simulation time is calculated by op_sim_time KP. By reception of each data packet, if the value of $\bar{L}_{sj}(t)$ for corresponding service class is more than the allocated buffer length for that service class, then $\bar{L}_{sj}(t)$ will not be updated and the packet is destroyed. Equation (3) is updated by the arrival of each probing packet for three service classes and the values of $B_{sj}$ are written in the probing packets. As the probing packet passes each router, the new values of $B_{sj}$, are compared to the stored values of previous router and finally the minimum values of $B_{sj}$ are found. Then, the summations of $B_{sj}$ values for three service classes are written in two acknowledgment packets according to equation (4). In order to perform load balancing, equation (5) is updated in router2, router0 and router3 in data transfer phase. The assumed maximum buffer length of EF, AF and BE classes are respectively 24, 53 and 374 packets.

Distinguishing the advantages of the load balancing algorithm, simulation results of two scenarios are compared to each other: The proposed algorithm is implemented in the first scenario without performing the load balancing algorithm and there is only a single path for the accepted traffic. The second scenario is similar to the first scenario except that the accepted traffic will be divided between two paths by load balancing algorithm.

In order to show the difference in the performance of the two scenarios, two parameters are evaluated: End-to-end delay is the time difference between sending a packet from the source and receiving it at the destination calculated in seconds. Throughput is the number of received packets at the destination in a pre-defined period calculated in packets per second.

The probing packets with the length of 50 bytes are sent through the network every 0.5 seconds and data packets will be sent after the probing phase with the length of 500 bytes. The proportion of the guaranteed rate of each service class to the output link capacity is the normalized WFQ weights which is 0.22, 0.33 and 0.44 respectively for EF, AF and BE classes. Main traffic uses a source with a constant rate of 4Mbps. As it is clear in the simulation results of Fig. 2 and Fig. 3, the performance of these two scenarios are the same.

The reason is that the source traffic rate is lower than the links' capacity and hence no packet losses will occur in the two scenarios. In order to show the difference between them the arrival traffic rate should be increased or network capacity should be decreased. In the next experiment, the effect of service rate degradation is evaluated on two scenarios. Since router1 is the most crowded point in the network, by decreasing its service rate, the advantages of the proposed algorithm will become clear. Simulation results are shown in Fig. 4 and Fig. 5.

In this case, by decreasing the service rate of router1, the average buffer lengths increase and congestion will occur in the first scenario. Therefore, packets queued in the buffers should wait longer times to be served in the routers. But in the second scenario since two paths exist, data packets have an alternative not to experience delay and loss as the first scenario. Due to non-deterministic network characteristics, it is not precise to compare the difference of two scenarios numerically. This experiment is repeated with a self-similar traffic and the simulation results are shown in Fig. 6 and Fig. 7.

The obtained simulation results are in accordance with the applied topology. In order to generalize the simulation results for higher scale networks, one should expand the applied topology to set more routers and source-destination pairs. Also number of selected path for load balancing can be increased for further investigations. The proposed network topology and the arrival rate of main traffic are the other selective parameters. Although it is predicted that changing these parameters lead to the same results, doing more simulation can improve the accuracy of the results.

This algorithm was implemented in one single DiffServ domain. How to extend it for inter domain usage is still needs more research. On the other hand, passing periodic probing packets through the network, limits the scalability of the proposed algorithm. Considering the effect of probing packets' overhead on simulation results is another research issue. In addition, an appropriate packet reordering mechanism can be added to the algorithm to make it more applicable for TCP streams. Implementing this algorithm on MPLS networks and wireless networks and considering the required changes may lead to a new research issues in NGN networks [11].

**Traffic Generation Parameters:**
Start Time (second): constant (5)
ON State Time (second): pareto (40, 1.4)
OFF State Time (Second): pareto (1, 1.4)
**Packet Generation Arguments:**
Interarrival Time (second): exponential (0.001)
Packet Size (byte): constant (500)

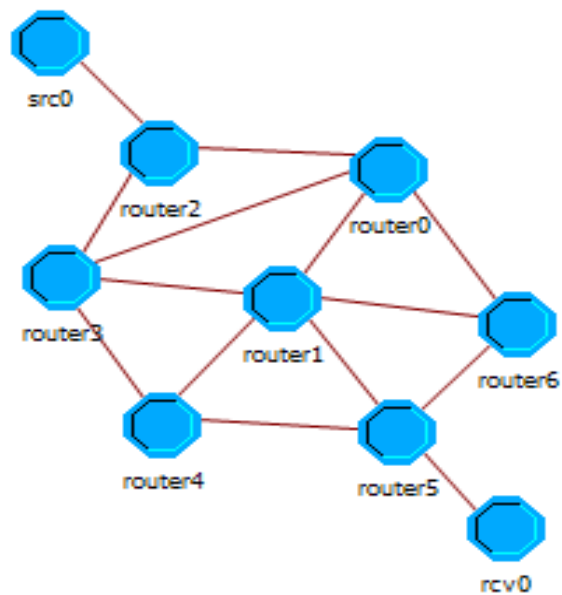

**Figure 1.** Applied topology

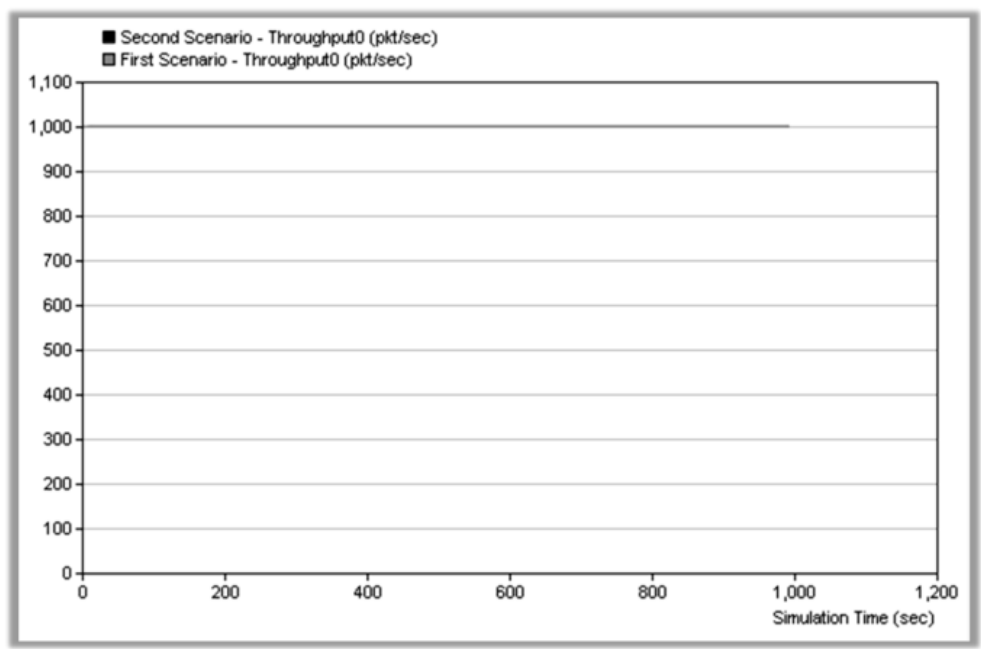

**Figure 2.** Average throughput of the two scenarios

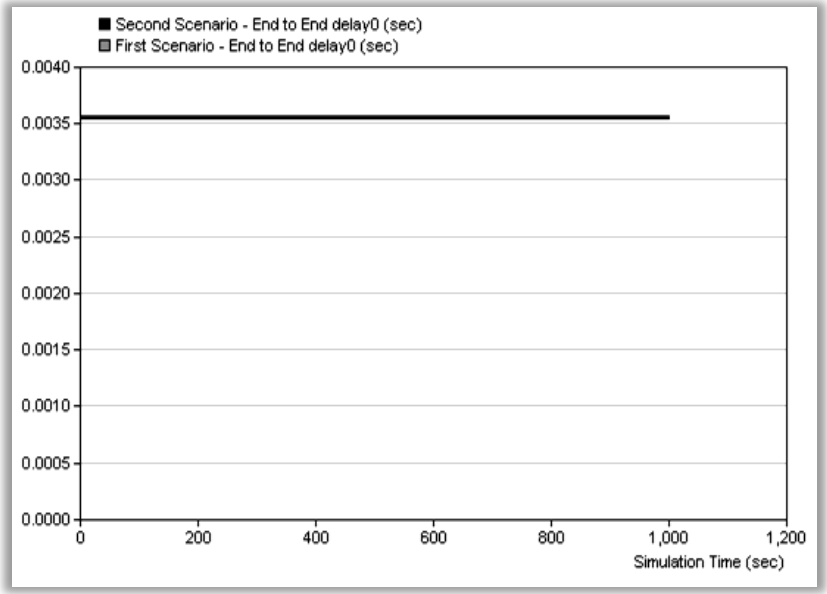

**Figure 3.** Average end-to-end delay of the two scenarios

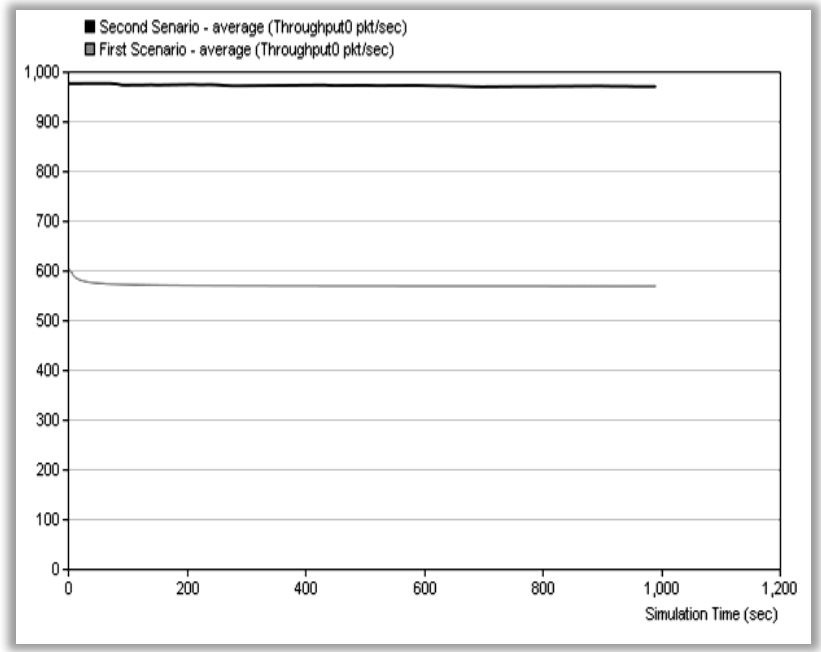

**Figure 4.** Average throughput of the two scenarios with decreased network capacity

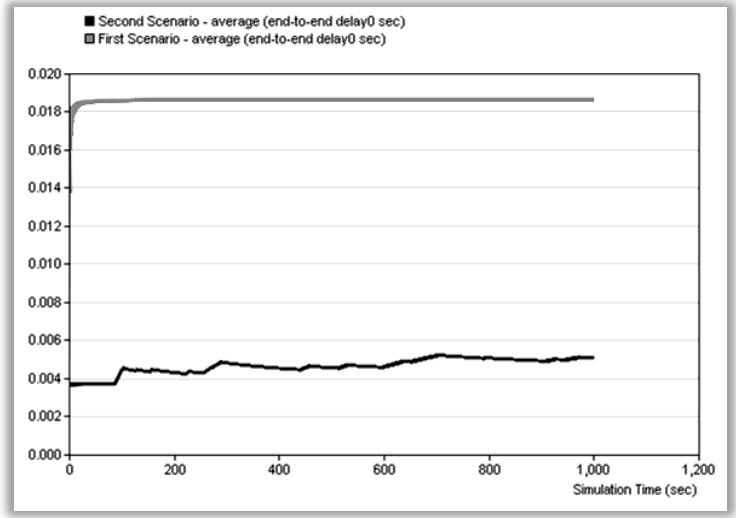

**Figure 5.** Average end-to-end delay of the two scenarios with decreased network capacity

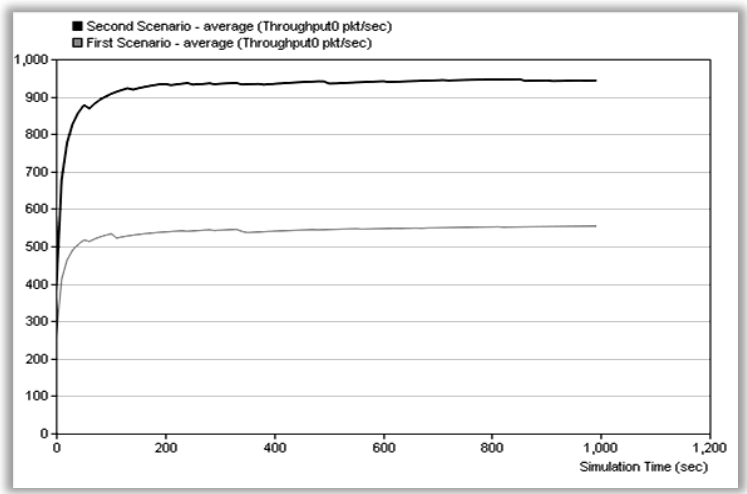

**Figure 6.** Average throughput of the two scenarios with decreased network capacity and self similar source

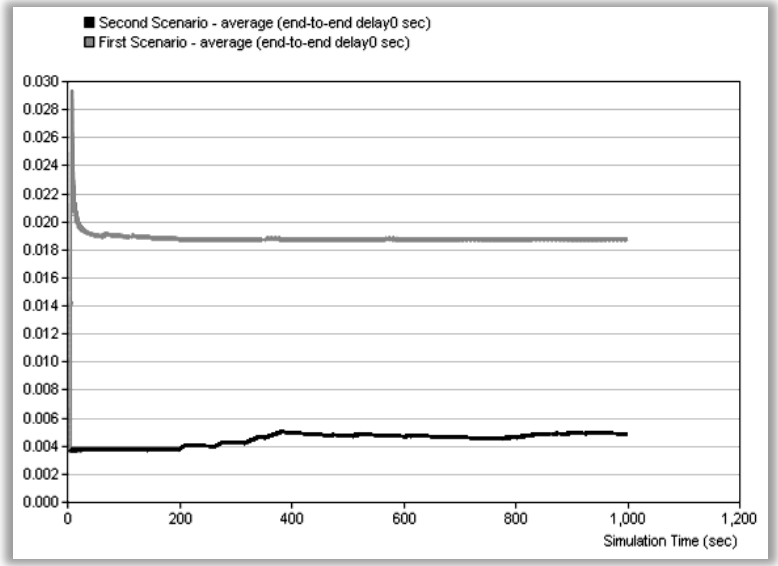

Figure 7. Average end-to-end delay of the two scenarios with decreased network capacity and self similar source

## 4. Conclusion

In this paper an EEAC algorithm was used to implement a new adaptive load balancing algorithm. The probing phase information represented a good estimation of network congestion. A heuristic equation is used for load balancing algorithm according to the obtained congestion information which is simple and scalable. Additionally, buffer length parameter played a key role in splitting traffic efficiently on multiple paths. Finding the most appropriate paths and splitting traffic among them, resulted in higher network throughput, better network utilization and lower end-to-end delay.

## References

[1] S. Upadhyaya, G. Devi, "Exploring Issues Related to Multipath Routing", Proceedings of the 5th National Conference, INDIACom, 2011.
[2] D. Awduche, A. Chiu, A. Elwalid, I. Widjaja, X. Xiao, "Overview and principles of internet traffic engineering", RFC 3272, 2002.
[3] Z. Cao, Z. Wang, E. Zegura, "Performance of hashing-based schemes for internet load balancing", IEEE INFOCOM, pp.332-341, 2000.
[4] C. Villamizar, "OSPF optimized multipath (OSPF-OMP)", Work in Progress, 1999.
[5] D. Thaler, C. Hopps, "Multipath issues in unicast and multicast next-hop selection", RFC 2991, 2000.
[6] H.S. Palakurthi, "Study of multipath routing for QoS provisioning", EECS, 2001.
[7] G. Yuan, Y. Chen, Y. Wei, S. Nie, "A distributable traffic-based MPLS dynamic load balancing scheme", Asia-Pacific Conference on Communications, pp. 684-689, 2005.
[8] D. Gao, Y. Shu, S. Liu, O.W.W. Yang, "Delay-based adaptive load balancing in MPLS networks", IEEE International Conference on Communications, vol. 2, pp. 1184-1188, 2002.

[9] X. He, H. Tang, M. Zhu, Q. Chu, "Flow-level based adaptive load balancing in MPLS networks", ChinaCOM, pp. 1-6, 2009.

[10] N. Moghim, S.M. Safavi, M.R. Hashemi, "Evaluation of a new end-to-end quality of service algorithm in differentiated service networks", IET Communications, vol. 4, no.14, pp. 1687-1695, 2010.

[11] Z. Vali, M. Hashemi, N. Moghim, "Proposing a load balancing algorithm with the help of an endpoint admission control algorithm to improve traffic engineering", JSCSE Journal, Advance Academic Publication, vol. 2, no. 6, pp.42-55, 2012.

[12] E. Oki, R. Rojas-Cessa, M. Tatipamula, C. Vogt, IP Quality Of Service, Advanced Internet Protocols, Services, and Applications, John Wiley & Sons, USA, 2012.

[13] P. Narvaez, K.Y. Siu, H.Y. Tzeng, "Efficient algorithms for multi-path link state routing", ISCOM'99, 1999.

[14] J. Yang, J. Ye, S. Papavassiliou, N. Ansari, "A flexible and distributed architecture for adaptive end-to-end QoS provisioning in next-generation networks", IEEE Journal On Selected Areas In Communications, vol. 23, no. 2, 2005.

[15] OPNET Modeler, Version 14.5.A